\documentclass[12pt,leqno]{article}
\usepackage{bbm}
\usepackage{amsfonts}
\usepackage{mathrsfs}
\usepackage {amssymb}
\usepackage {amsmath}
\usepackage{amsthm}
\usepackage{latexsym}
\usepackage{indentfirst}
\usepackage{enumitem}
\usepackage{bm}

\newtheorem{thm}{Theorem}

\newcommand{\pf}{{\bf Proof. \ }}
\def\dse#1{\vskip 0.6cm\noindent
        {\large\bf #1}
        \vskip 0.4cm}
\usepackage{lastpage}
\usepackage{epsfig}
\usepackage{leftidx}

\usepackage{anysize}
\marginsize{3cm}{3cm}{2cm}{2cm} \normalsize

\begin{document}
\title{Good Self-dual Generalized Quasi-Cyclic Codes Exist}

\author{MinJia Shi\thanks{School of Mathematical Sciences, Anhui University, Anhui $230601$, China (E-mail: smjwcl.good@163.com)}, Yan Liu and Patrick Sol\'e\thanks{CNRS/LTCI, Telecom ParisTech, Universit\'e Paris-Saclay, 75013 Paris}}
\date{}
\maketitle

\begin{abstract}
We show that there are good long binary generalized quasi-cyclic self-dual (either Type I or Type II) codes.
\end{abstract}

\noindent\textbf{Keywords}: Cubic construction; Gilbert-Varshamov bound; generalized quasi-cyclic codes; self-dual codes.

\noindent{\bf{{AMS(2010) 94A15, 94B20, 94B60}}}

\section{Introduction}

It has been known for more than forty years that good long self-dual codes exist \cite{MST}, and for more than thirty years that there are good long quasi-cyclic codes of rate ${1}/{2}$ \cite{K}. Only fifteen years ago, it was proved that good long self-dual quasi-cyclic codes exist \cite{LS2}. More recently, the class of Generalized Quasi-Cyclic Codes was introduced in \cite{S}, and studied further in \cite{EY}.

In this note, we show that good long self-dual generalized quasi-cyclic codes exist. Building on well-known mass formulas for self-dual binary and self-dual codes over $\mathbb{F}_{16}$, we derive a modified Gilbert-Varshamov bound for long binary self-dual generalized quasi-cyclic codes.

The proof uses the cubic and quintic constructions of \cite{LS1}, \cite{DNS} and the proof technique of \cite{LS2}.

\section{Known facts and notations} \label{sec:1}

\textbf{Definition 2.1.} Let $m_1,m_2,\cdots,m_{\ell}$ be postive integers and set $R_j=\mathbb{F}_q[x]/(x^{m_j}-1).$ An $\mathbb{F}_q[x]$ submodule of $R'=R_i\times \cdots\times R_\ell$ is called a {\bf generalized quasi-cyclic code} or shortly a GQC code of index $(m_1, m_2, \cdots, m_{\ell})$. The $m_i$'s are called the {\em co-indices}.\\

Note that if $C$ is a GQC code of length $(m_1, m_2, \cdots, m_{\ell})$ with $m=m_1=m_2=\cdots=m_{\ell})$, then $C$ is a quasi-cyclic code with length $m{\ell}$. Further if ${\ell}=1$, then $C$ is a cyclic code of length $m$. \\

We assume that all binary codes are equipped with the Euclidean inner product and all the $\mathbb{F}_{16}$-codes are equipped with the Hermitian inner product. The latter condition is necessary, when using the cubic construction and quintic construction, to ensure that the resulting binary code is Euclidean self-dual. Self-duality in the following discussion is with respect to these respective inner products. A binary self-dual code is said to be of Type II if and only if all its weights are multiples of 4 and of Type I otherwise. We first recall some background
material on mass formulas for self-dual binary and $16$-ary codes \cite[Chap. 12]{NRS}. \\

\textbf{Proposition 2.2.} Let $\ell$ be an even positive integer.

(1) The number of self-dual binary codes of length $\ell$ is given by

$$N(2,\ell)=\prod_{i=1}^{\frac{\ell}{2}-1}(2^i+1).$$

(2) Let $v$ be a codeword of length $\ell$ and even Hamming weight, other than $\textbf{0}$ and $\textbf{1}$. The number of self-dual binary codes of length $\ell$ containing $\textbf{v}$ is given by $$M(2,\ell)=\prod_{i=1}^{\frac{\ell}{2}-2}(2^i+1).$$

(3) The number of self-dual $\mathbb{F}_{16}$ codes of length $\ell$ is given by

$$N(16,\ell)=\prod_{i=1}^{\frac{\ell}{2}-1}\frac{(2^{4i+2}+1)}{12\times 5^{\ell} \times \ell!}.$$

(4) Let $v$ be a codeword of length $\ell$ and even Hamming weight, other than $\textbf{0}$ and $\textbf{1}$. The number of self-dual binary codes of length $\ell$ containing $\textbf{v}$ is given by $$M(16,\ell)=\prod_{i=1}^{\frac{\ell}{2}-2}\frac{(2^{4i+2}+1)}{12\times 5^{\ell} \times \ell!}.$$

\textbf{Proposition 2.3.} Let $\ell$ be an positive integer divisible by 8.

(1) The number of Type II binary codes of length $\ell$ is given by
$$T(2,\ell)=2\prod_{i=1}^{\frac{\ell}{2}-2}(2^i+1).$$

(2) Let $v$ be a codeword of length $\ell$ and Hamming weight divisible by 4, other than $\textbf{0}$ and $\textbf{1}$. The number of Type II binary codes of length $\ell$ containing $\textbf{v}$ is given by $$S(2,\ell)=2\prod_{i=1}^{\frac{\ell}{2}-3}(2^i+1).$$

\section{Combinatorial bounds} \label{sec:1}
\subsection{Cubic construction}
Let $C_1$ denote a binary code of length $3m_1$ and $C_2$ a $\mathbb{F}_{16}$-code of length $5m_2$. We construct a binary code $C$ of length $3m_1$ by the cubic construction [3]. Define a map

$$\Phi: C_1 \times C_2 \longrightarrow F_2^{3m_1}$$ by the rule

$$\Phi(x, \textbf{a}+\textbf{b}\omega):= (x+\textbf{a},x+\textbf{b},x+\textbf{a}+\textbf{b})$$ where $\textbf{a},\textbf{b}$ are binary vectors of length $m_1$, and we write $\mathbb{F}_4$. Then we can define the code $C$ as Im $(\Phi)$

$$C:=\{\Phi(x, \textbf{a}+\textbf{b}\omega)|x\in C_1, \textbf{a}+\textbf{b}\omega\in C_2\}.$$ In [2], the author proved that $C$ is a $m_1$ quasi-cyclic code and $C$ is self-dual if and only if both $C_1$ and $C_2$ are, and $C$ is of Type II if and only if $C_1$ is of Type II and $C_2$ is self-dual.
\subsection{Quintic construction}
 We construct a binary code $C$ of length $5m_2$ by the quintic construction [3]. Define a map

$$\Phi: C_1 \times C_2 \longrightarrow F_2^{5m_2}$$ by the rule

$\Phi(x, \textbf{a}_0+\textbf{a}_1\alpha+\textbf{a}_2\alpha^2+\textbf{a}_3\alpha^3):= (x+\textbf{a}_0, x+\textbf{a}_0+\textbf{a}_1, x+\textbf{a}_1+\textbf{a}_2, x+\textbf{a}_2+\textbf{a}_3, x+\textbf{a}_3)$
where $\textbf{a}_i'$s are vectors of length $m_2$ over $\mathbb{F}_{16}$ such that $\sum_{i=0}^{3}a_{i}\alpha^{i}\in C_2'$ are binary vectors of length $m_2$,

It is easy to check that $C$ is a quasi-cyclic code and $C$ is self-dual if and only if both $C_1$ and $C_2$ are, and $C$ is of Type II if and only if $C_1$ is of Type II and $C_2$ is self-dual.

We assume henceforth that $C$ is a self-dual code constructed in
the above way. Any codeword $c$ in $C$ must necessarily have even Hamming
weight. Suppose that $c$ corresponds to the pair $(c_1, c_2)$, where $c_1\in C_1$ and $c_2\in C_2$. Since both $C_1$ and $c_2$ are self-dual, it follows that
$c_1$ and $c_2$ must both have even Hamming weights. When $c\neq 0$, there are three possibilities for the pair $(c_1, c_2)$:\\

\begin{enumerate}
\item $c_1\neq 0, c_2\neq 0$
\item $c_1=0, c_2\neq 0$
\item $c_1\neq0, c_2=0$
\end{enumerate}
We try to enumerate the number of words $\mathbf{c}$ in each of these categories for a given weight $d$ ($d$ even).\\

{\bf Convention:} ${M \choose N}=0$ if $N$ is not an integer.\\
Denote by $A_j(\ell,d)$ the number of words of type $j=1,2,3.$ For type (i) we only give the upper bound
$$ A_1(\ell,d)\le {{5 \ell} \choose d}-A_2(\ell,d)-A_3(\ell,d).$$
For type (ii), if the Hamming weight $d$ is even we use Corollary 3.2 of \cite{LS2} to show
$$A_2(\ell,d)\le {\ell \choose {d/2}} 15^{d/2}.$$
For type (iii), similarly, we have
$$A_3(\ell,d)\le {\ell \choose {d/5}}.$$
Combining these observations with the counting functions of the preceding sections we see that the number of self-dual binary 5-quasi-cyclic codes of length $5\ell$ whose
minimum weight is $< d$ is bounded above by

$$ A_1(\ell,d)M(2,\ell) M(16,\ell)+A_2(\ell,d)N(2,\ell)M(16,\ell)+ A_2(\ell,d)M(2,\ell)N(16,\ell).$$
We are now in a position to give a sufficient existence condition for self-dual $5$-QC codes of length $5\ell,$ and distance $\ge d.$
{\thm If the inequality $$\sum_{e<d}{{5 \ell} \choose e}+2^\frac{\ell-2}{2} \sum_{e<d}{\ell \choose {e/2}} 15^{e/2}+2^{2\ell-2}\sum_{e<d}{\ell \choose {e/5}}< (2^\frac{\ell-2}{2}+1)(2^{2\ell-2}+1)$$ holds then there is a self-dual $5$-QC code of length $5\ell,$ and distance $\ge d.$}

\pf Since the total number of self-dual $5-$QC code of length $5\ell$ is $N(2,\ell)N(16,\ell),$ there will be at least on such code of distance $\ge d$ if
$$ A_1(\ell,d)M(2,\ell) M(16,\ell)+A_2(\ell,d)N(2,\ell)M(16,\ell)+ A_2(\ell,d)M(2,\ell)N(16,\ell)<N(2,\ell)N(16,\ell).$$
Dividing both sides by $M(2,\ell) M(16,\ell)$ the above condition can be re written as
$$ A_1(\ell,d)+A_2(\ell,d)\frac{N(2,\ell)}{M(2,\ell) }+ A_2(\ell,d)\frac{N(16,\ell)}{ M(16,\ell)}<\frac{N(2,\ell)N(16,\ell)}{M(2,\ell) M(16,\ell)}.$$
Combining the above estimates for the $A_i(\ell,d)$ with the expressions for $M(2,\ell),N(2,\ell),N(16,\ell),M(16,\ell)$ of the previous section, the result follows.
\qed

The analogous result for Type II codes is as follows.
{\thm Let $\ell$ be a multiple of $8.$ If the inequality $$\sum_{e<d}{{5 \ell} \choose e}+2^\frac{\ell-4}{2} \sum_{e<d}{\ell \choose {e/2}} 15^{e/2}+2^{2\ell-2}\sum_{e<d}{\ell \choose {e/5}}< (2^\frac{\ell-4}{2}+1)(2^{2\ell-2}+1)$$ holds then there is a doubly even self-dual $5$-QC code of length $5\ell,$ and distance $\ge d.$}

\pf
In the previous argument replace $N(2,\ell)$ by $T(2,\ell)$ and $M(2,\ell)$ by $S(2,\ell).$
\qed

\section{Asymptotic bounds}
We will require the $q-$ary entropy function
defined for $0<x<\frac{q-1}{q}.$
$$H_q(x)=x\log_q(q-1)-x\log_q x-(1-x)\log_q(1-x).$$
Note for future use the identity
\begin{equation}\label{ident}
4H_{16}(x)=x\log_2(15)+H_2(x).
\end{equation}
Denote by $V_q(n,r)=\sum_{j=o}^r(q-1)^j{n\choose j},$ the volume of the Hamming ball of radius $r.$
By Lemma 2.10.3 of \cite{HP} we have
\begin{equation}\label{entrop}
\lim_{n \rightarrow \infty}\frac{1}{n}\log_qV_q(n,\lfloor x n\rfloor)=H_q(x).
\end{equation}
\subsection{Cubic codes}
We summarize the asymptotic results of \cite{LS2} in the following theorem

{\thm \label{3QC}
There exist infinite families of self-dual binary
\begin{itemize}
\item Type I  codes $C_3^I$
\item Type II codes $C_3^{II}$
\end{itemize}
$3$-QC of length $3\ell$ and of relative asymptotic distance $\delta \ge H_2(1/2)\approx 0.11.$}
\subsection{Quintic codes}
We give an analogue of the cubic codes results as follows.
{\thm \label{5QC}
There exist infinite families of self-dual binary
\begin{itemize}
\item Type I  codes $C_5^I$
\item Type II codes $C_5^{II}$
\end{itemize}
QC codes of length $5\ell$ and  of relative asymptotic distance $\delta \ge H_2(1/2)\approx 0.11.$}

\pf We only sketch the proof for the Type I case, the Type II case being similar. Assume $\ell$ large and $d\sim 5\ell \delta.$ We compare the RHS of Theorem 1 which is, up to subexponential factors, of the order of $2^{5\ell/2}$ to the three terms of the LHS. The first and the second term, applying equation \ref{entrop} for $q=2,$ yield the classical condition $H_2(\delta)=1/2.$
The second term, after applying equation \ref{entrop} for $q=16,$ combined with identity \ref{ident}  yield the transcendental equation
$$8=5\delta \log_2(15)+H_2(5\delta), $$
whose unique solution can be seen numerically to be $< 0.11.$
\qed
\subsection{GQC codes}
We are ready for the main result of this note.
{\thm
There exist infinite families of self-dual binary
\begin{itemize}
\item Type I  codes $G^I$
\item Type II codes $G^{II}$
\end{itemize}
GQC codes of length $8\ell$ and with $\ell$ co-indices 3 and $\ell$ co-indices 5 and of relative asymptotic distance $\delta \ge \frac{3H_2(1/2)}{8}\approx 0.041.$}

\pf
The family $G^I$ is obtain by taking the direct sum of the codes in $C_3^I$ and $C_5^I$ corresponding to the same value of $\ell.$
The family $G^{II}$ is obtain by taking the direct sum of the codes in $C_3^{II}$ and $C_5^{II}$ corresponding to the same value of $\ell.$
\qed

\dse{Acknowledgments}

This research is
supported by National Natural Science Foundation of China (61202068) and Technology Foundation for Selected Overseas Chinese Scholar, Ministry of Personnel of China (05015133).

\par

\end{document}